%% file: qubiq_main.tex
\documentclass[review,3p,authoryear]{elsarticle}
\UseRawInputEncoding
\usepackage[multiple]{footmisc}
\usepackage{pdflscape}
\usepackage{amsmath}

\usepackage{enumitem}
\usepackage{rotating} %{figuresright}
\usepackage{lineno}
\usepackage{amssymb}
\usepackage{sidecap}
\usepackage{graphicx}
\usepackage{caption}
\usepackage{amsmath}
\usepackage{bbm}
\usepackage{subcaption}
\usepackage{multirow}
\usepackage{wrapfig}
\usepackage{xcolor}
\newcommand\crule[3][black]{\textcolor[HTML]{#1}{\rule{#2}{#3}}}
\usepackage{lipsum}
\usepackage{nicefrac}
\usepackage{ctable}
\usepackage{booktabs,tabularx}
\usepackage{adjustbox}
\usepackage{floatrow}
\usepackage{color}
\usepackage{colortbl}
\usepackage{float}
\restylefloat{table}
\newcolumntype{x}{>{\raggedright\arraybackslash}X}
\usepackage{bbding}
\usepackage{amsmath}
\usepackage{romannum}
\usepackage[colorlinks=true,
            linkcolor=magenta,
            urlcolor=gray,
            citecolor=cyan]{hyperref}
\usepackage{slashbox}
\usepackage{arydshln}
\newfloatcommand{capbtabbox}{table}[][\FBwidth]
\usepackage{mathtools}
\usepackage{blindtext}
\usepackage{enumitem}
\usepackage{algpseudocode}
\usepackage{xcolor}
\usepackage{adjustbox}
\usepackage{array}
\usepackage[linesnumbered,ruled,vlined]{algorithm2e}
\usepackage[final]{pdfpages}
\usepackage{amssymb}% http://ctan.org/pkg/amssymb
\usepackage{pifont}% http://ctan.org/pkg/pifont
\newcommand{\cmark}{\ding{51}}%
\newcommand{\xmark}{\ding{55}}%
\usepackage{multicol}   
\usepackage{lipsum} % dummy text for the MWE
\newcounter{includepdfpage}
\newcounter{currentpagecounter}

\newcolumntype{t}{>{\hsize=0.25\hsize}x}
\newcolumntype{s}{>{\hsize=0.75\hsize}x}
\newcolumntype{X}{>{\hsize=1.5\hsize}x}
\newcolumntype{R}[2]{%
    >{\adjustbox{angle=#1,lap=\width-(#2)}\bgroup}%
    l%
    <{\egroup}%
}

\newcommand{\tabincell}[2]{\begin{tabular}{@{}#1@{}}#2\end{tabular}}

\usepackage{xparse}
\NewDocumentCommand{\rot}{O{90} O{1em} m}{\makebox[#2][l]{\rotatebox{#1}{#3}}}%

\modulolinenumbers[5]
\newpageafter{author}

\begin{document}
\begin{frontmatter}
\title{QUBIQ: Uncertainty Quantification for Biomedical Image Segmentation Challenge}

\input{authors}

\begin{abstract}
Uncertainty in medical image segmentation tasks, especially inter-rater variability, arising from differences in interpretations and annotations by various experts, presents a significant challenge in achieving consistent and reliable image segmentation. This variability not only reflects the inherent complexity and subjective nature of medical image interpretation but also directly impacts the development and evaluation of automated segmentation algorithms. Accurately modeling and quantifying this variability is essential for enhancing the robustness and clinical applicability of these algorithms. 
We report the set-up and summarize the benchmark results of the Quantification of Uncertainties in Biomedical Image Quantification Challenge (QUBIQ), which was organized in conjunction with International Conferences on Medical Image Computing and Computer-Assisted Intervention (MICCAI) 2020 and 2021. The challenge focuses on the uncertainty quantification of medical image segmentation which considers the omnipresence of inter-rater variability in imaging datasets. The large collection of images with multi-rater annotations features various modalities such as MRI and CT; various organs such as the brain, prostate, kidney, and pancreas; and different image dimensions 2D-vs-3D. A total of 24 teams submitted different solutions to the problem, combining various baseline models, Bayesian neural networks, and ensemble model techniques. The obtained results indicate the importance of the ensemble models, as well as the need for further research to develop efficient 3D methods for uncertainty quantification methods in 3D segmentation tasks.

\end{abstract}

% \begin{keyword}
% spine, vertebrae, segmentation, labelling, computed tomography
% \end{keyword}
\end{frontmatter}

\section{Introduction}
\begin{figure}[h]
  \centering
    \includegraphics[width=0.8\linewidth]{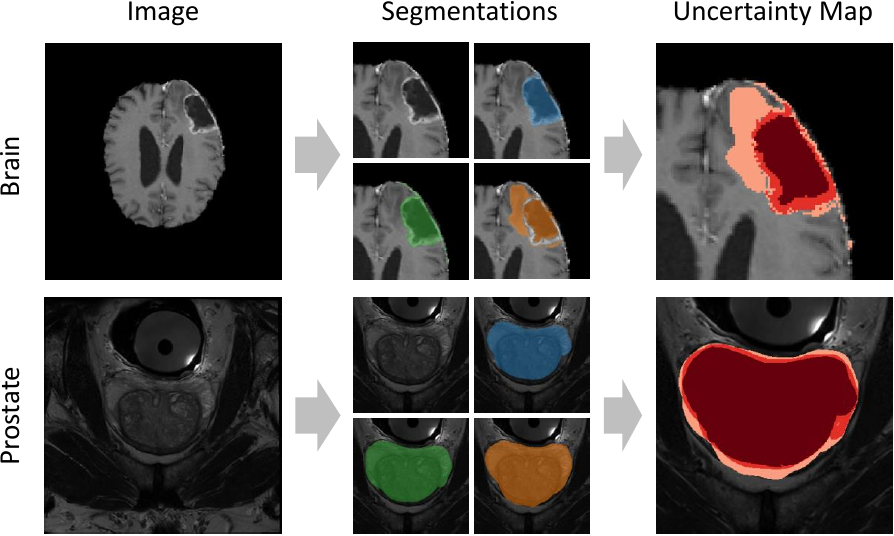}\\
    \caption{\small{Visualisation of the multi-rater segmentation masks on brain and prostate MRI scans and their derived uncertainty map.}}
  \label{fig:sample_image}
\end{figure}

\paragraph{Background}
The segmentation of anatomical structures and pathologies in medical images frequently encounters substantial inter-rater variability \citep{lazarus2006bi,watadani2013interobserver}, which in turn significantly impacts downstream supervised-learning tasks and clinical decision-making processes. This variability becomes especially pronounced in the context of medical imaging, where manual annotations are often limited and costly to acquire \citep{kofler2023approaching}. A notable example of this challenge is the segmentation of liver lesions in CT scans, which is inherently complex even for experienced experts, due to the variability in lesion location, contrast, and size among different patients \citep{joskowicz2019inter}. It has been observed that the range of variability in manual delineations for various structures and observers is extensive, encompassing a wide spectrum of structures and pathologies, as shown in Figure \ref{fig:sample_image}. The involvement of only two or three observers may be inadequate to capture the full breadth of potential variability in the outlines of the targeted structures. This variability, intrinsic to the biological problem, the imaging modality, and the expertise of the annotators has not yet been adequately addressed in the design of computerized algorithms for medical image quantification \citep{kofler2021we}.

\paragraph{Uncertainty quantification}
Current methods for modeling uncertainty in predicted image segmentations primarily stem from general statistical model considerations, ensemble approaches involving resampling of training datasets, and aggregating multiple segmentation results, or systematic modifications to the predictive algorithm, as seen in techniques like Monte Carlo (MC) dropout. Yet, the exact delineation of segmented structures within an image inherently carries uncertainty, which is both task-specific and dependent on the dataset. Importantly, this uncertainty can be directly extrapolated from annotations made by multiple human experts. To our knowledge, there are currently no datasets available specifically for evaluating the accuracy of probabilistic model predictions against such multi-expert ground truths. Furthermore, there is a lack of consensus on which uncertainty quantification procedures yield realistic estimates and which do not. 

\paragraph{Objective}
The primary goal of the challenge is to establish a benchmark for algorithms that generate uncertainty estimates (such as probability scores and variability regions) in medical imaging segmentation tasks. The focus is to compare these algorithmic outputs against the uncertainties ascribed by human annotators in the local delineation of structures across various biomedical imaging segmentation tasks. These tasks include, but are not limited to, the segmentation of lesions (such as brain, pancreas, or prostate tumors) and anatomical structures (like brain, kidney, prostate, and pancreas). Multiple expert annotations have been gathered for several CT and MR image datasets to quantify boundary delineation variability.

\paragraph{Contributions} 
In an effort to assess the latest methods in uncertainty quantification for medical image segmentation, we organized the Uncertainty Quantification of Biomedical Image Quantification Challenge (QUBIQ) at MICCAI-2020 and MICCAI-2021. This paper highlights three major contributions to this field. Firstly, we introduce a new, publicly available multi-rater, multi-center, multi-modality dataset that includes both 2D and 3D segmentation tasks. Secondly, we present the setup and summarize the findings of our QUBIQ uncertainty quantification benchmarks held at two grand challenges. Lastly, we review, evaluate, rank, and analyze the state-of-the-art algorithms that emerged from these benchmarks.

%%%%%%%%% %

\section{Prior Work on Approaches and Datasets}
\label{sec:priorwork}
\color{black}

\subsection{Prior work.} There is a body of literature that models uncertainty and inter-rater variability in biomedical image segmentation \citep{le2016sampling,sabuncu2010generative,kwon2020uncertainty,roy2019bayesian,ilg2018uncertainty}. Some of the prior methods directly extract uncertainty estimates from trained models, either by augmenting the input image \cite{wang2019aleatoric} or by generating multiple potential segmentations using MC dropout \cite{nair2020exploring}. Others modify techniques into ensemble methods that generate multiple parallel predictions \cite{ilg2018uncertainty} or by running multiple models in parallel \cite{calisto2020adaen}. \cite{kofler2021robust} extend this further to create an ensemble of multiple approaches from the literature and create a system to alert the user if there is low segmentation agreement within the ensemble.
In contrast, others explicitly model inter-rater uncertainty. In Probabilistic U-Net, Kohl et al. \cite{kohl2018probabilistic} use variational inference to learn a prior distribution of variability, from which they sample plausible segmentations, while Baumgartner et al. \cite{baumgartner2019phiseg} extend this to a hierarchical model capable of modeling uncertainty at different levels of abstraction within the U-Net architecture. Monteiro et al. \cite{monteiro2020stochastic} explicitly model uncertainty by learning a low-rank pixel-wise covariance matrix.

\subsection{Publicly available datasets.} 
Table \ref{tab:datasets} showcases available datasets for uncertainty quantification task. Most of the datasets feature multi-rater labeling. Each focuses on a particular pathological or healthy anatomy segmentation task. Therefore, the datasets contain either contain 2D or 3D images, CT or MRI modality. The QUBIQ challenge offers a dataset composed of multiple tasks for both image dimensions and imaging modalities.

%%%%%Public datasets%%%%
\begin{table}[h]
\tiny
\resizebox{\textwidth}{!}{
\begin{tabular}{llcccllllc}
\toprule
Dataset & Modality & Target & 2D & 3D & \#Images &  multi-rater  \\ 
\midrule
%\citep{clark2013cancer,
LIDC-IDRI \citep{armato2011lung}& CT & Lung nodule & \cmark & \textcolor{orange}{\xmark} & 1,018 & \cmark \\
MICCAI-2012 \citep{litjens2012pattern}  & MRI & Prostate & \cmark & \textcolor{orange}{\xmark}  & 48 & \cmark  \\
ISBI-2015 \citep{styner20083d} & MRI & MS lesion  & \textcolor{orange}{\xmark} & \cmark & 21 & \cmark  \\
BraTS \citep{mehta2020uncertainty}  & MRI & brain tumor  & \textcolor{orange}{\xmark} & \cmark & 335 &  \textcolor{orange}{\xmark}\\
\textbf{QUBIQ}  & CT,MRI  & \textbf{six tasks} & \cmark & \cmark   & **  & \cmark \\
\bottomrule
\end{tabular}}
\caption{Overview of publicly available medical datasets for uncertainty quantification in image segmentation tasks. (to be updated)}
\label{tab:datasets}
\end{table}
%%%%%%%%% %

% \subsection{Approaches for uncertainty quantification in medical image segmentation} 
% lala
% \paragraph{Deep learning-based image segmentation}

% \paragraph{Methods based on MC dropout}

% \paragraph{Methods based on Bayesian neural network}

\section{QUBIQ challenge}

\subsection{QUBIQ datasets}
% Number of images?
\subsubsection{Dataset creation.}
For the adult glioma segmentation task, we employ three label sets.
The first label set is the original label from the BraTS adult glioma segmentation challenge \citep{bakas2019identifying}.
% [TODO describe human rater -> reference brats]
Additionally, we use two algorithm-based labels obtained from BraTS Toolkit \citep{kofler2020brats}.
To generate these, we first generate five algorithmic 
\citep{isensee2019no,mckinley2019ensembles,feng2020brain,feng2020brain,zhao2019multi,mckinley2020triplanar} glioma segmentations.
Subsequently, we fuse these using basic majority voting and SIMPLE fusion \citep{langerak2010label}.

% https://bitbucket.org/neuronflow/qubiq_brats_seg/src/master/src/fusion.py

%%%%%Qubiq datasets%%%%
\begin{table}[H]
\tiny
\resizebox{\textwidth}{!}{
\begin{tabular}{llcccrrl}
\toprule
Dataset & Modality & [2020,2021] & 2D & 3D & \#Images & \#Tasks & Source (NEED to double check \\ 
\midrule
%\citep{clark2013cancer,
Prostate segmentation & MRI & [\cmark,\cmark] & \cmark & \textcolor{orange}{\xmark} & 55 & 2 & ETH Zürich \\
Brain growth segmentation & MRI &  [\cmark,\cmark] & \cmark & \textcolor{orange}{\xmark} & 39 & 1 & University of Zürich  \\
Brain tumor segmentation & multimodal MRI &  [\cmark,\cmark]  &\cmark & \textcolor{orange}{\xmark} & 32 & 3 & University of Pennsylvania \\
Kidney segmentation  & CT &  [\cmark,\cmark]  & \cmark & \textcolor{orange}{\xmark} & 24 & 1 & Technical University of Munich\\
Pancreas segmentation & CT &  [\textcolor{orange}{\xmark},\cmark]  & \textcolor{orange}{\xmark} & \cmark & 38 & 1 & University of Pennsylvania \\
Pancreatic lesion segmentation  & CT &  [\textcolor{orange}{\xmark},\cmark]  & \textcolor{orange}{\xmark} & \cmark & 21 & 1 & University of Pennsylvania \\
\bottomrule
\end{tabular}}
\caption{Overview of QUBIQ datasets and the sub-tasks}
\label{tab:qubic}
\end{table}
%%%%%%%%% %
% \paragraph{Contributors} 

% \paragraph{Data diversity}

% \paragraph{Annotation protocol} 

\subsection{Evaluation metrics and ranking}

%\subsubsection{Metrics}

For the evaluation, each participant had to segment the given binary structures and predict the distribution of the experts' labels by returning one mask with continuous values between 0 and 1 which is supposed to reproduce the average segmentations of the experts.    

Predictions and continuous ground truth labels are compared by thresholding the continuous labels at predefined thresholds and calculating the volumetric overlap of the resulting binary volumes using the Dice score (the continuous ground truth labels are obtained by averaging multiple experts' annotations). To this end, both the ground truth and prediction are binarized at ten probability levels (0.1, 0.2, ..., 0.8, 0.9). Dice scores for all thresholds are averaged. 

% Dice scores is averaged across all tasks and all image data sets. The participant performing best according to this average was named the "winner" of the challenge. 

The Q-Dice, a staged Dice score, is used to quantify the quality of the predicted probability map $p$ against the ground truth $y$ in $L$ discrete probability levels, formulated as:

\begin{equation}
    T_L(p, l) = 
    \begin{dcases}
        \mathbbm{1}\left\{\frac{l}{L} \leq p < \frac{l + 1}{L}\right\}, & \text{if } 0 \leq l < L - 1 \\
        \mathbbm{1}\left\{\frac{l}{L} \leq p \leq \frac{l + 1}{L}\right\}, & \text{if } l = L - 1 \\
    \end{dcases}
\end{equation}

Compared to the original Dice score, Q-score quantifies the uncertainty by comparing the prediction and ground truth maps at different confidence levels. Since in most cases experts agree on most parts of the annotations, the variance of different Q-score demonstrates how well the prediction modeled the uncertainty on the borders of the structure of interest.

\label{sec:Dice_score}

% \paragraph{Generalized Energy Distance}
% \label{sec:GED}

%\subsubsection{Ranking method} (Fernando)
% What platform did we use?

\subsection{Challenge events}
\begin{figure}[h]
  \centering
    \includegraphics[width=0.7\linewidth]{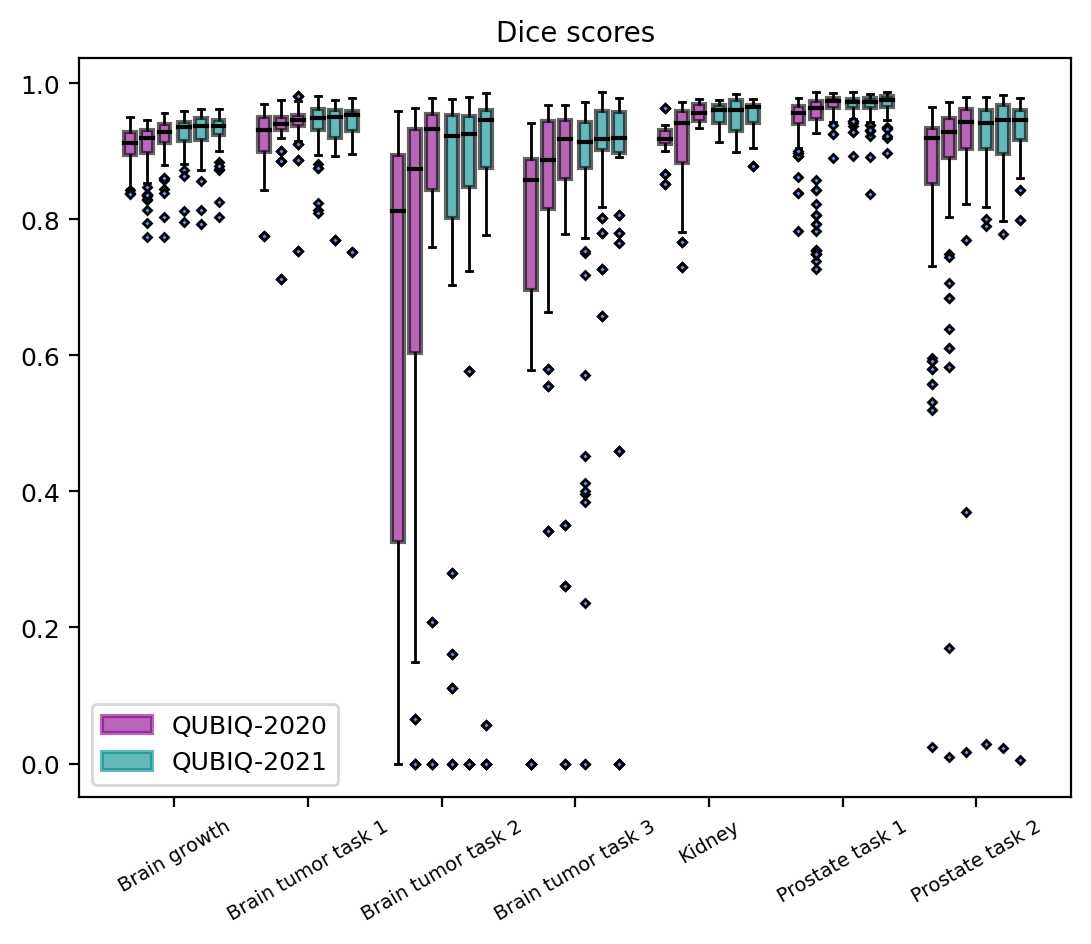}\\
    \caption{\small{Pictorial evaluation of QUBIQ challenge for different tasks over two years. Observe that for every task, the top-performing methods produce higher scores in 2021 than in 2020. Also, the methods in 2021 are more competitive than in 2020.}}
  \label{fig:overview}
\end{figure}

The QUBIQ challenge was organized within the MICCAI conference using the Grand Challenge platform. Below in Tables 3, 4, and 5, we provide descriptions of algorithms across the two iterations of the QUBIQ challenge (QUBIQ2020 and QUBIQ2021). Fig. \ref{fig:overview} quantitatively compares the algorithms over the two iterations of the challenge.

%\paragraph{MICCAI-2020}

%(Bran)

%\paragraph{MICCAI-2021}

%(Bran)

%%%%%%%%%%%%%%%%%%%%%%%%%%%%%%%%%

%%%%%%%%%%%%%%%%%%%%%%%%%

%\end{center}

%%%%%%%%%%%%%%%%%%%%%%%%%%%%%

%\subsection{Online evaluation platform}
% What platform did we use?

%%%%%%%%%%%%%%%%%%%%%%%%%%%%%%%%%%%%%%%%%%%%%%%%%%%%%%%%%%%%%%%%%%%

%\section{Participating Methods}
%\label{sec:methods}

%(mentioned the table, comments)

% \paragraph{Algorithms and architectures} 

% \paragraph{Critical components of the uncertainty quantification methods} 

% \paragraph{Post-processing} 

% \paragraph{Features of top-performing methods} 

%%%%%%%%%%%%%%%%%%%%%%%%%%%%%%%%%%%%%%%%%%%%%%%%%%%%%%%%%%%%%%%%%%%

\subsection{Results}
\label{sec:results}
In Tables \ref{tab:quibiq2020_results} and \ref{tab:quibiq2021_results}, we show the leaderboard for both iterations of the challenge.

%\subsection{MICCAI-2020}

%Here the results from MICCAI 2020 QUIBIQ challenge

% \usepackage{graphicx}
\begin{table}[h!]
\resizebox{\textwidth}{!}{%
\begin{tabular}{llllllllll}
\specialrule{.1em}{0em}{-.1em}
\textbf{Ref. Name} &
  \textbf{Brain-growth} &
  \textbf{Brain-tumor} &
  \textbf{Brain-tumor} &
  \textbf{Brain-tumor} &
  \textbf{Kidney} &
  \textbf{Prostate} &
  \textbf{Prostate} &
  \textbf{Average} &
  \textbf{Average} \\
\textbf{} &
  \textbf{} &
  \textbf{Task 1} &
  \textbf{Task 2} &
  \textbf{Task 3} &
  \textbf{} &
  \textbf{Task 1} &
  \textbf{Task 2} &
  \textbf{Ranking} &
  \textbf{Dice} \\
  \specialrule{.05em}{-0.1em}{0em}
\crule[b0b0ff]{0.15cm}{0.15cm} \textbf{Jun\_Ma}               & \cellcolor{blue!10}\textbf{0.921} & \cellcolor{blue!10}\textbf{0.936} & \cellcolor{blue!10}\textbf{0.809} & \cellcolor{blue!10}0.822 & 0.310 & \cellcolor{blue!10}\textbf{0.970} & \cellcolor{blue!10}\textbf{0.918} & \cellcolor{blue!10}\textbf{7.857} & \cellcolor{blue!10}0.812 \\
\crule[550500]{0.15cm}{0.15cm} \textbf{Yanwu\_Yang}           & 0.893 & 0.917 & \cellcolor{blue!10}0.699 & \cellcolor{blue!10}\textbf{0.836} & 0.825 & 0.937 & \cellcolor{blue!10}0.878 & \cellcolor{blue!10}7.143 & \cellcolor{blue!10}\textbf{0.855} \\
\textbf{Macaroon}              & 0.878 & 0.848 & 0.528 & 0.690 & 0.238 & 0.937 & 0.890 & \cellcolor{blue!10}5     & 0.715 \\
\crule[baba00]{0.15cm}{0.15cm} \textbf{Raghavendra\_Selvan 2} & 0.885 & 0.899 & \cellcolor{blue!10}0.617 & 0.682 & 0.695 & 0.883 & 0.800 & 4.857 & 0.780 \\
\crule[baba00]{0.15cm}{0.15cm} \textbf{Raghavendra\_Selvan}   & \cellcolor{blue!10}0.907 & 0.874 & 0.602 & 0.690 & 0.639 & 0.858 & 0.780 & 4.714 & 0.764 \\
\crule[00b0b0]{0.15cm}{0.15cm} \textbf{Wei\_Ji}               & \cellcolor{blue!10}0.900 & 0.755 & 0.323 & 0.605 & \cellcolor{blue!10}0.915 & \cellcolor{blue!10}0.941 & 0.845 & 4.714 & 0.755 \\
\crule[00ff00]{0.15cm}{0.15cm} \textbf{Xiang\_Li}             & 0.865 & \cellcolor{blue!10}0.931 & 0.513 & 0.556 & \cellcolor{blue!10}0.903 & 0.914 & 0.872 & 4.714 & \cellcolor{blue!10}0.793 \\
\crule[008900]{0.15cm}{0.15cm} \textbf{Ujjwal\_Baid}          & 0.840 & 0.782 & 0.406 & 0.568 & \cellcolor{blue!10}\textbf{0.956} & 0.891 & 0.702 & 3.143 & 0.735 \\
\textbf{Maykol\_Campos}        & 0.849 & 0.799 & 0.522 & 0.613 & 0.805 & 0.838 & 0.630 & 2.857 & 0.722 \\
\textbf{anysys99}              & 0.818 & 0.893 & 0.485 & \cellcolor{blue!10}0.724 & -     & 0.890 & 0.804 & -     & -     \\
\crule[00b000]{0.15cm}{0.15cm} \textbf{Davood\_Karimi}        & 0.874 & \cellcolor{blue!10}0.900 & 0.452 & -     & 0.785 & \cellcolor{blue!10}0.947 & \cellcolor{blue!10}0.897 & -     & - \\
\specialrule{.1em}{0em}{-.1em}
\end{tabular}%
}
\caption{
Results QUIBIQ 2020 ordered according to the ranking score.  The top 3 performing teams are highlighted in blue color.  Notice that only teams participating in all tasks are considered for the overall ranking.
}
\label{tab:quibiq2020_results}

\end{table}

\begin{table}[h!]
\resizebox{\textwidth}{!}{%
\begin{tabular}{llllllllllll}
\specialrule{.1em}{0em}{-.1em}
\textbf{Team} &
  \textbf{Brain-growth} &
  \textbf{Brain-tumor} &
  \textbf{Brain-tumor} &
  \textbf{Brain-tumor} &
  \textbf{Kidney} &
  \textbf{Prostate} &
  \textbf{Prostate} &
  \textbf{Pancreas} &
  \textbf{Pancreatic} &
  \textbf{Average} &
  \textbf{Average} \\
\textbf{} &
  \textbf{} &
  \textbf{Task 1} &
  \textbf{Task 2} &
  \textbf{Task 3} &
  \textbf{} &
  \textbf{Task 1} &
  \textbf{Task 2} &
  \textbf{} &
  \textbf{Lesion} &
  \textbf{Ranking} &
  \textbf{Dice} \\
  \specialrule{.05em}{-0.1em}{0em}
\textbf{Peng-Cheng\_Shi}       & 0.929 & \cellcolor{blue!10}0.938 & \cellcolor{blue!10}0.819 & \cellcolor{blue!10}0.847 & \cellcolor{blue!10}\textbf{0.954} & \cellcolor{blue!10}0.969 & \cellcolor{blue!10}0.920 & 0.550 & \cellcolor{blue!10}0.272 & \cellcolor{blue!10}\textbf{11.111} & \cellcolor{blue!10}\textbf{0.800} \\
\crule[ff0000]{0.15cm}{0.15cm} \textbf{Yingbin\_Bai}          & 0.915 & 0.928 & \cellcolor{blue!10}0.793 & 0.815 & 0.940 & 0.968 & \cellcolor{blue!10}0.920 & \cellcolor{blue!10}0.579 & 0.205 & \cellcolor{blue!10}9.333  & \cellcolor{blue!10}0.785 \\
\crule[550500]{0.15cm}{0.15cm} \textbf{Lara\_Dular}           & \cellcolor{blue!10}0.928 & \cellcolor{blue!10}0.938 & \cellcolor{blue!10}\textbf{0.820} & \cellcolor{blue!10}\textbf{0.899} & 0.467 & 0.958 & 0.909 & 0.499 & \cellcolor{blue!10}0.283 & \cellcolor{blue!10}8.778  & 0.745 \\
\crule[550500]{0.15cm}{0.15cm} \textbf{Lawrence\_Schobs}      & 0.300 & \cellcolor{blue!10}0.939 & 0.780 & 0.798 & 0.503 & \cellcolor{blue!10}0.969 & 0.915 & \cellcolor{blue!10}\textbf{0.683} & \cellcolor{blue!10}\textbf{0.330} & 8.556  & 0.691 \\
\crule[00b0b0]{0.15cm}{0.15cm}\textbf{Sabrican\_Cetindag}    & \cellcolor{blue!10}0.928 & 0.932 & 0.769 & \cellcolor{blue!10}0.883 & 0.839 & 0.964 & \cellcolor{blue!10}0.922 & 0.409 & 0.231 & 8.444  & 0.764 \\
\crule[eeb000]{0.15cm}{0.15cm} \textbf{Yucong\_Chen}          & 0.916 & 0.927 & 0.775 & 0.840 & \cellcolor{blue!10}0.952 & 0.952 & 0.907 & 0.572 & 0.130 & 7.889  & \cellcolor{blue!10}0.775 \\
\crule[e4e400]{0.15cm}{0.15cm} \textbf{Hoang\_Long\_Le}       & 0.912 & 0.899 & 0.680 & 0.754 & 0.706 & \cellcolor{blue!10}\textbf{0.971} & \cellcolor{blue!10}\textbf{0.927} & \cellcolor{blue!10}0.575 & 0.246 & 7.444  & 0.741 \\
\crule[00b0ee]{0.15cm}{0.15cm}\textbf{Dewen\_Zeng}           & 0.927 & \cellcolor{blue!10}\textbf{0.940} & 0.695 & 0.835 & 0.894 & 0.947 & 0.911 & 0.423 & 0.126 & 7.111  & 0.744 \\
\crule[00ff00]{0.15cm}{0.15cm}\textbf{Joao\_Lourenco\_Silva} & \cellcolor{blue!10}\textbf{0.931} & 0.929 & 0.750 & 0.797 & 0.511 & 0.968 & \cellcolor{blue!10}0.920 & 0.075 & 0.068 & 6.333  & 0.661 \\
\crule[00b000]{0.15cm}{0.15cm} \textbf{Anindo\_Saha}          & 0.892 & 0.917 & 0.695 & 0.740 & \cellcolor{blue!10}0.950 & 0.936 & 0.859 & 0.546 & 0.194 & 5.222  & 0.748 \\
\crule[888700]{0.15cm}{0.15cm} \textbf{Wang\_Xiong}           & 0.893 & 0.905 & 0.589 & 0.784 & 0.930 & 0.916 & 0.862 & 0.557 & 0.204 & 5.222  & 0.738 \\
\crule[baba00]{0.15cm}{0.15cm} \textbf{Ishaan\_Rajesh}        & 0.892 & 0.919 & 0.638 & 0.704 & 0.858 & 0.861 & 0.799 & 0.316 & 0.122 & 3.111  & 0.679 \\
\crule[00b0aa]{0.15cm}{0.15cm} \textbf{Stephan\_Huschauer}    & 0.719 & 0.865 & 0.525 & 0.551 & 0.856 & 0.911 & 0.842 & 0.423 & 0.118 & 2.444  & 0.646 \\
\crule[008900]{0.15cm}{0.15cm} \textbf{Jiachen\_Zhao}         & 0.873 & 0.844 & 0.547 & 0.787 & 0.835 & 0.931 & 0.884 & -     & -     & -      & -     \\
\crule[545400]{0.15cm}{0.15cm} \textbf{Jimut\_Bahan\_Pal}     & 0.869 & 0.842 & 0.456 & 0.690 & 0.769 & 0.833 & 0.781 & -     & -     & -      & -     \\
\crule[eeee77]{0.15cm}{0.15cm}\textbf{Mohammad\_Eslami}      & 0.848 & 0.404 & 0.377 & 0.236 & 0.716 & 0.883 & 0.816 & -     & -     & -      & -     \\
\crule[888700]{0.15cm}{0.15cm} \textbf{Shengbo\_Gao}          & 0.802 & 0.885 & 0.627 & 0.661 & 0.910 & -     & -     & 0.557 & 0.130 & -      & -     \\
\crule[eeeeee]{0.15cm}{0.15cm}\textbf{Xiaofeng\_Liu}         & 0.800 & -     & -     & -     & -     & -     & -     & -     & -     & -      & -     \\
\crule[00b000]{0.15cm}{0.15cm}\textbf{Timothy\_S}            & 0.780 & -     & -     & -     & -     & -     & -     & -     & -     & -      & -\\
\specialrule{.1em}{0em}{-.1em}

\end{tabular}%
}
\caption{
Results QUIBIQ 2021 ordered according to the ranking score.  The top 3 performing teams are highlighted in blue color.  Notice that only teams participating in all tasks are considered for the overall ranking.
}
\label{tab:quibiq2021_results}
\end{table}

\section{Conclusion}
\label{sec:conclusion}

In this paper, we report on the results of the QUBIQ challenge (Quantification of Uncertainties in Biomedical Image Quantification Challenge), which was organized in conjunction with International Conferences on Medical Image Computing and Computer-Assisted Intervention (MICCAI). Quantifying uncertainty in medical imaging is paramount for image analysis, as inter-rater variability is omnipresent in imaging datasets. Such quantification could reduce barriers to adopting learnable algorithms into clinical practice. With the QUBIQ challenge, we aim to fill the empty space among the medical imaging challenges, which are dominated by competition in deterministic segmentation, ignoring the importance of uncertainty prediction.

\section*{Acknowledgement}
The research is supported through the SFB 824,
subproject B12, as well as by Deutsche Forschungsgemeinschaft (DFG) via TUM International Graduate School of Science and Engineering (IGSSE), GSC 81.
We acknowledge support by the Helmut Horten Foundation and by the Translational Brain Imaging Training Network (TRABIT) under the EU ‘Horizon 2020’ research \& innovation program (Grant agreement ID: 765148).
Research reported in this publication was partly supported by the National Institutes of Health (NIH) under award numbers NIH/NCI:U01CA242871 and
NIH/NINDS:R01NS042645.

\begin{center}
\begin{landscape}
\begin{table}
\tiny
\caption{Details of the participating teams’ methods in QUBIQ-Challenge-2020.} 
\setlength\tabcolsep{4pt} % default value: 6pt
\begin{tabular}{l l l l l l l}
\toprule
\tabincell{l}{\textbf{Lead Author} \& \\ \textbf{Team Members}} & \tabincell{l}{\textbf{Method, Architecture} \& \\ \textbf{Modifications}} & \tabincell{l}{\textbf{Data Augmentation}} & \textbf{Loss Function}  & \textbf{Pre-processing}  &\textbf{Label Processing} & \textbf{Ensemble strategy}\\
\midrule

\tabincell{l}{\crule[b0b0ff]{0.15cm}{0.15cm} \textbf{Jun Ma}}  & \tabincell{l}{Multiple 2D U-Nets\\(one per annotator).}  &  \tabincell{l}{None}  & \tabincell{l}{Cross-entropy \\\& Dice loss} & \tabincell{l}{None} & \tabincell{l}{None} & \tabincell{l}{Averaging}\\
\midrule

\tabincell{l}{\crule[00b000]{0.15cm}{0.15cm} \textbf{Davood Karimi}}  & \tabincell{l}{2D U-Net with additional\\connections between coarse\\and fine feature layers in the\\encoder. Dynamic loss\\weighting for harder classes.\\ Multi-task training approach.}  &  \tabincell{l}{None}  & 1-Dice similarity loss & \tabincell{l}{Zero mean, unit variance\\standardization.} & \tabincell{l}{Averaging \\annotations} & \tabincell{l}{None}\\
\midrule

\tabincell{l}{\crule[e4e400]{0.15cm}{0.15cm} \textbf{Ming Feng}; \\ Kele Xu, \\Yin Wang}  & \tabincell{l}{2D U-Net trained with\\ground truth \& predictions\\ binarized at different levels,\\ averaging Dice score of\\each prediction.}  &  \tabincell{l}{Random scaling \\ $[0.9, 1.1]$}  & \tabincell{l}{Weighted cross-entropy \\ \& Dice loss} & \tabincell{l}{Resizing to \\ 256 $\times$ 256 (brain tumor)\\512 $\times$ 512 (kidney \& prostrate).\\Normalization to [0, 255].} & \tabincell{l}{Averaging \\annotations} & None\\
\midrule

\tabincell{l}{\crule[baba00]{0.15cm}{0.15cm} \textbf{Raghavendra Selvan}}  & \tabincell{l}{Multi-channel U-Net,\\one channel for each rater.\\Based on concept of \\ Normalizing Flows.}  &  \tabincell{l}{None}  & \tabincell{l}{Planar Flow \\ \& Dice loss}  & \tabincell{l}{None} & \tabincell{l}{None} & \tabincell{l}{None}\\
\midrule

\tabincell{l}{\crule[008900]{0.15cm}{0.15cm} \textbf{Ujjwal Baid}; \\ Prasad Dutande, \\Shubham Innani, \\ Bhakti Baheti, \\Sanjay
Talbar}  & \tabincell{l}{ResNet34 based encoder\\-decoder. Different annotations\\included as individual\\copies in training set.}  &  \tabincell{l}{Rotation, flip \\\& scaling}  & None & \tabincell{l}{Resizing to\\256 $\times$256 (brain)\\512 $\times$512 (kidney)\\640 $\times$640 (prostate).} & \tabincell{l}{None} & \tabincell{l}{None}\\
\midrule

\tabincell{l}{\crule[545400]{0.15cm}{0.15cm} \textbf{Wesam Adel}; \\ Mustafa A. Elattar}  & \tabincell{l}{2D U-Net trained\\with averaged annotations\\and as a regression problem.}  &  \tabincell{l}{None}  & Weighted KL-divergence & \tabincell{l}{Resizing brain to \\256 $\times$ 256 with\\ rotation and elastic\\deformation.} & \tabincell{l}{None} & \tabincell{l}{None}\\
\midrule

\tabincell{l}{\crule[00ff00]{0.15cm}{0.15cm} \textbf{Xiang Li}}  & \tabincell{l}{U Net with attention,\\4x downsampling for Kidney\\and 5x for others.}  &  \tabincell{l}{None}  & Weighted cross-entropy& \tabincell{l}{Cropping to \\128 $\times$ 128 (kidney)\\416 $\times$ 416 (prostate)} & \tabincell{l}{Averaging \\annotations} & \tabincell{l}{None}\\
\midrule

\tabincell{l}{\crule[550500]{0.15cm}{0.15cm} \textbf{Yanwu Yang}; \\ Ting Ma}  & \tabincell{l}{2D U-Net with multiple \\branches. Instance Norm \\ instead of Batch Norm.\\One model per annotation\\integrated using auxiliary loss.}  &  \tabincell{l}{None}  & \tabincell{l}{Cross-entropy \\\& Dice loss} & \tabincell{l}{MRI: z-score normalization.\\CT: centering on ROI\\and rescaling to $[0, 1]$} & \tabincell{l}{None} & \tabincell{l}{Averaging}\\
\midrule

\tabincell{l}{\crule[00b0b0]{0.15cm}{0.15cm} \textbf{Wei Ji}; \\ Wenting Chen\\Shuang Yu\\Kai Ma\\Li Cheng\\Linlin
Shen\\Yefeng Zheng}  & \tabincell{l}{U-Net with Resnet-34 \\encoder. One output \\channel per label and \\one model per annotation \\ integrated using auxiliary loss.}  &  \tabincell{l}{None}  & Cross-entropy & \tabincell{l}{Resizing to \\512 $\times $512} & \tabincell{l}{Both fused final \& individual\\ labels. Combining labels via \\ averaging, random sampling \\\& label sampling.} & \tabincell{l}{Weighted average}\\

\bottomrule
%    \midrule
\end{tabular}
\label{tab:method_1}
\end{table}
% \caption{Details of the participating teams’ methods.}

\end{landscape}
\end{center}
% \clearpage

%%%%%%%%%%%%%%%%%%%%%%%%%%%%%

%%%%%%%%%%%%%%%%%%
%\begin{center}
\begin{landscape}
\begin{table}
\tiny
\caption{Details of the participating teams’ methods in QUBIQ-Challenge-2021 (part 1).}  
\setlength\tabcolsep{4pt} % default value: 6pt

\begin{tabular}{l l l l l l l}
\toprule
\tabincell{l}{\textbf{Lead Author} \& \\ \textbf{Team Members}} & \tabincell{l}{\textbf{Method, Architecture} \& \\ \textbf{Modifications}} & \tabincell{l}{\textbf{Data Augmentation}} & \textbf{Loss Function}  & \textbf{Pre-processing}  &\textbf{Label Processing} & \textbf{Ensemble strategy}\\
\midrule

\tabincell{l}{\crule[00b000]{0.15cm}{0.15cm}  \textbf{Anindo Saha};\\ Henkjan Huisman}  & \tabincell{l}{Probabilistic U-Net\\with MC Dropout.}  & \tabincell{l}{Gaussian noise, horizontal\\flip, rotation, translation\\\& scaling} & \tabincell{l}{KL-Divergence\\\& Dice loss}  & \tabincell{l}{z-Score normalization,\\ centre cropping to\\ 512 $\times$512 (Kidney)\\256$\times$256 (Brain)\\640$\times$640 (Prostrate)}& None & \tabincell{l}{Deep ensemble \\ (averaging)}\\[-2pt]
\midrule

\tabincell{l}{\crule[e4e400]{0.15cm}{0.15cm} \textbf{Hoang Long Le}; \\Jin Tae Kwak}  & \tabincell{l}{DeepLabv3 \& EfficientNet \\ (latter for classifying \\ pancreas existence). 9 binary\\ ground truths from thresholding.} & \tabincell{l}{Gaussian noise, horizontal \\\& vertical flip, rotation, shift\\scaling, blur random brightness}& Dice loss & \tabincell{l}{Normalization to [0, 255]} &\tabincell{l}{Averaging \\annotations} & \tabincell{l}{Multiplying binary\\segmentation map with\\threshold value \& taking \\ per pixel maximum.} \\ [-2pt]
\midrule

\tabincell{l}{\crule[baba00]{0.15cm}{0.15cm} \textbf{Ishaan Bhat};\\Hugo J. Kuijf } & \tabincell{l}{Probabilistic U-Net \\with MC Dropout.}   & \tabincell{l}{Random flip,\\rotation, brightness \& \\contrast}&\tabincell{l}{Cross-entropy \&\\ KL-Divergence} & \tabincell{l}{z-Score normalization,\\Resizing to 256$\times$256 (brain)\\512$\times$512 (kidney)\\512$\times$512 (pancreas)} & \tabincell{l}{None} & \tabincell{l}{Deep ensemble\\(averaging)} \\ [-2pt]
\midrule

\tabincell{l}{\crule[008900]{0.15cm}{0.15cm} \textbf{Jiachen Zhao}} & \tabincell{l}{U-Net}   & \tabincell{l}{Random flip}&\tabincell{l}{Dice loss} &  \tabincell{l}{Resizing to\\256 $\times$ 256} & \tabincell{l}{Averaging \\annotations} & \tabincell{l}{None} \\ [-2pt]
\midrule

\tabincell{l}{\crule[545400]{0.15cm}{0.15cm} \textbf{Jimut Bahan Pal}} & \tabincell{l}{Multiple U-Nets\\(one per annotation).}   & \tabincell{l}{None}&\tabincell{l}{Focal Tversky} & \tabincell{l}{None} & \tabincell{l}{None} & \tabincell{l}{None} \\ [-2pt]
\midrule

\tabincell{l}{\crule[00ff00]{0.15cm}{0.15cm} \textbf{Jo\~ao Louren\c{c}o Silva};\\Arlindo L. Oliveira} & \tabincell{l}{U-Net with\\ EfficientNet-B0 \\encoder.}   & \tabincell{l}{Rotation, horizontal\\\& vertical flip,\\translation \& zoom}&\tabincell{l}{Cross-entropy} & \tabincell{l}{None} & \tabincell{l}{Averaging \\ annotations} & \tabincell{l}{None} \\ [-2pt]
\midrule

\tabincell{l}{\crule[550500]{0.15cm}{0.15cm} \textbf{Lawrence Schobs}} & \tabincell{l}{Multiple nnU-Nets\\(one per annotator,\\2D for pancreas, 3D for other).} & \tabincell{l}{Gaussian noise,\\ rotation, scaling\\mirroring \& inhomogeneity}&\tabincell{l}{Dice loss} & \tabincell{l}{Resizing prostate images\\to $640 \times 640$. Image\\ sampling and\\normalization.} & \tabincell{l}{Averaging\\ annotations} & \tabincell{l}{None} \\ [-2pt]
\midrule

\tabincell{l}{\crule[550500]{0.15cm}{0.15cm} \textbf{Martin \~{Z}ukovec};\\Lara Dular\\\~{Z}iga \~{S}piclin} & \tabincell{l}{nnU-Net. Multi-task \\ training approach with labels\\ 0--N (N annotators + background).}   & \tabincell{l}{Gaussian noise,\\ rotation, scaling\\mirroring \& inhomogeneity}&\tabincell{l}{Dice loss} & \tabincell{l}{Image sampling\\and normalization} & \tabincell{l}{Addition of\\ segmentations} & \tabincell{l}{None} \\ [-2pt]
\midrule

\tabincell{l}{\crule[00b0b0]{0.15cm}{0.15cm} \textbf{Sabri Can Cetindag};\\Mert Yergin\\Deniz Alis \\Ilkay Oksuz} & \tabincell{l}{nnU-Net, training one U-Net\\per annotator, then adding\\segmentation map output\\as extra channels.}   & \tabincell{l}{None}&\tabincell{l}{Cross-entropy \\ \& Dice loss} & \tabincell{l}{None} & \tabincell{l}{Average of\\ annotations\\for stage 2} & \tabincell{l}{None} \\ [-2pt]
\midrule

\tabincell{l}{\crule[888700]{0.15cm}{0.15cm} \textbf{Xiong Wang};\\Shengbo Gao\\Weifeng Hu\\Xuan Pei} & \tabincell{l}{2D: U-Net (one per annotator,\\ obtained via label fusion),\\ MaskNet, and DeepLab V3+.\\3D: SegResNet.} & \tabincell{l}{None}&\tabincell{l}{Weighted focal\\ \& Dice loss} & \tabincell{l}{None} & \tabincell{l}{Label fusion\\for training\\individual models} & \tabincell{l}{Weighted combination} \\ [-2pt]
\midrule

\tabincell{l}{\crule[ff00ff]{0.15cm}{0.15cm} \textbf{YingLin Zhang};\\Wei Wang\\Ruiling Xi\\Lingxi Zeng\\Huiyan Lin} & \tabincell{l}{UNet, UNet++,\\ and TransUNet.}   & \tabincell{l}{Gaussian noise, horizontal\\ \& vertical flip, rotation,\\translation, zoom, brightness,\\ sharpness, contrast, blur \\\& elastic deformation}&\tabincell{l}{Multi-level Dice loss} & \tabincell{l}{Center-cropping images \\ to ROI. Discarding images\\with unclear ROI} & \tabincell{l}{Majority voting,\\ cumulative division \& even\\ division  of segmentations} & \tabincell{l}{Weighted combination} \\ [-2pt]
\midrule

\tabincell{l}{\crule[ff0000]{0.15cm}{0.15cm} \textbf{Yingbin Bai};\\Maoying Qiao\\Dadong Wang\\Tongliang Liu} & \tabincell{l}{UNet++ with \\EfficientNet-B7 encoder.\\}   & \tabincell{l}{None}&\tabincell{l}{Multi-level Dice loss} & \tabincell{l}{None} & \tabincell{l}{Weighted combination\\of individual\\segmentation maps} & \tabincell{l}{None} \\ [-2pt]

\bottomrule
%    \midrule
\end{tabular}
\label{tab:method_2}
\end{table}
% \caption{Details of the participating teams’ methods.}
\end{landscape}
%\end{center}

%\begin{center}
\begin{landscape}
\begin{table}
\tiny
\caption{Details of the participating teams’ methods in QUBIQ-Challenge-2021 (part 2).}  
\setlength\tabcolsep{4pt} % default value: 6pt

\begin{tabular}{l l l l l l l}
\toprule
\tabincell{l}{\textbf{Lead Author} \& \\ \textbf{Team Members}} & \tabincell{l}{\textbf{Method, Architecture} \& \\ \textbf{Modifications}} & \tabincell{l}{\textbf{Data Augmentation}} & \textbf{Loss Function}  & \textbf{Pre-processing}  &\textbf{Label Processing} & \textbf{Ensemble strategy}\\
\midrule

\tabincell{l}{\crule[00b0ee]{0.15cm}{0.15cm}  \textbf{Dewen Zeng}; \\ Yukun Ding\\Yiyu Shi}  & \tabincell{l}{2D U-Net with multiple\\loss functions (one each\\for individual annotators,\\aggregated label \& multi-scale\\threshold).}  & \tabincell{l}{None} & \tabincell{l}{Cross-entropy}  & \tabincell{l}{z-Score normalization,\\ resizing} & \tabincell{l}{Averaging\\annotations} & \tabincell{l}{None}\\[-2pt]
\midrule

\tabincell{l}{\crule[00b000]{0.15cm}{0.15cm}  \textbf{Yanwu Yang}; \\ Xutao Guo\\Yiwei Pan\\Pengcheng Shi\\Haiyan Lv\\Ting Ma}  & \tabincell{l}{2D U-Net with one decoder\\per annotator and\\Layer Norm and skip-connections.}  & \tabincell{l}{None} & \tabincell{l}{Cross-entropy \& Dice loss\\on individual decoders\\A cross loss between\\different decoders \\\& an auxiliary loss between average \\prediction \& average ground truth}  & \tabincell{l}{z-Score normalization,\\ resizing} &\tabincell{l}{Averaging different\\labels} & \tabincell{l}{Deep ensemble \\ (averaging)}\\[-2pt]
\midrule

\tabincell{l}{\crule[00b0aa]{0.15cm}{0.15cm}  \textbf{Stephan Huschauer}}  & \tabincell{l}{High-Resolution Network (HRNet)\\ with stem layers replaced by\\ 2D wavelet scattering transformation.}  & \tabincell{l}{None} & \tabincell{l}{None}  & \tabincell{l}{Resizing to\\512 $\times$ 512}&\tabincell{l}{Averaging\\annotations} & \tabincell{l}{Deep ensemble \\ (averaging)}\\[-2pt]
\midrule

\tabincell{l}{\crule[eeeeee]{0.15cm}{0.15cm}  \textbf{Xiaofeng Liu};\\Fangxu Xing\\Georges El Fakhri\\Jonghye Woo}  & \tabincell{l}{Variational Inference encoding\\multi-annotator variability with a\\latent variable model.}  & \tabincell{l}{None} & \tabincell{l}{Cross-entropy \\\& L2 reconstruction loss}  & \tabincell{l}{None}&\tabincell{l}{Averaging\\annotations} & \tabincell{l}{Deep ensemble \\ (averaging)}\\[-2pt]
\midrule

\tabincell{l}{\crule[eeb000]{0.15cm}{0.15cm}  \textbf{Yucong Chen};\\Guanqi He\\Zhitong Gao\\Xuming He}  & \tabincell{l}{2D U-Net with multiple\\ decoders, one for each annotator.}  & \tabincell{l}{Random cropping (training),\\sliding window (inference)} & \tabincell{l}{None}  & \tabincell{l}{None}&\tabincell{l}{None} & \tabincell{l}{Deep ensemble \\ (averaging)}\\[-2pt]
\midrule

\tabincell{l}{\crule[eeee77]{0.15cm}{0.15cm}  \textbf{Mohammad Eslami};\\Farzin Soleymani\\Anirudh Ashok\\Bernd Bischl\\Mina Rezaei}  & \tabincell{l}{Uncertainty-aware \\progressive GAN. \\Encoder modeled using\\2D U-Net \& \\Patch Discriminators \\from Pix2Pix}  & \tabincell{l}{None} & \tabincell{l}{Multi-stage \\GAN loss and \\ Soft Dice loss}  & \tabincell{l}{Intensity values \\noramlized between\\0-255}&\tabincell{l}{Averaging \\annotations} & \tabincell{l}{None}\\[-2pt]
\midrule

\tabincell{l}{\crule[00b000]{0.15cm}{0.15cm}  \textbf{Timothy Sum Hon Mun};\\Simon J Doran\\Paul Huang\\Christina Messiou\\Matthew D Blackledge}  & \tabincell{l}{2D U-Net with \\Monte Carlo dropout.}  & \tabincell{l}{None} & \tabincell{l}{Dice loss}  & \tabincell{l}{None}&\tabincell{l}{None} & \tabincell{l}{None}\\[-2pt]
% \midrule

\bottomrule
%    \midrule
\end{tabular}
\label{tab:method_2}
\end{table}
% \caption{Details of the participating teams’ methods.}
\end{landscape}
 
%\section*{Acknowledgement}

\label{sec:summary}

%%%%%%%%%%%%%%%%%%%%%%%%%%%%%%%%%%%%%%
\clearpage
\bibliographystyle{./01_paper/model5-names}
\typeout{}
\bibliography{./01_paper/reference.bib}

%%%%%% Appendix

%\clearpage
%\newpage
\appendix

\end{document}

%% file: authors.tex
%%%%%%%%%%%%%%%%%%%%%%%%%%%
% senior authors who organize, analyze the results and write the manuscript %
%%%%%%%%%%%%%%%%%%%%%%%%%%%

\author[tum_info,martinos,uzh_dqbm]{Hongwei Bran~Li}
\author[tum_info,uzh_dqbm,deisar,tum_TranslaTUM]{Fernando~Navarro}
\author[tum_info,tum_TranslaTUM]{Ivan~Ezhov}
\author[tum_info]{Amirhossein~Bayat}
\author[tum_info,MIT]{Dhritiman~Das}
\author[hai,tum_info,translatum,tum_neuro]{Florian~Kofler}
\author[tum_info]{Suprosanna~Shit}
\author[tum_info,uzh_dqbm]{Diana~Waldmannstetter}
\author[tum_info]{Johannes~C.~Paetzold}
\author[tum_info]{Xiaobin~Hu}
\author[tum_neuro]{Benedikt Wiestler}
\author[uzh_dqbm]{Lucas~Zimmer}
\author[uzh_dqbm]{Tamaz~Amiranashvili}
\author[uzh_dqbm]{Chinmay~Prabhakar}
\author[tum_info]{Christoph~Berger}
\author[tum_info,tum_TranslaTUM]{Jonas~Weidner}

%%%%%%%%%%%%%%%%%%%%%%%%%%%
% co-organizers (and data contributors) %
%%%%%%%%%%%%%%%%%%%%%%%%%%%

%%% data contributors, not sorted
\author[RadOnc_UPenn]{Michelle~Alonso-Basanta}
\author[RadOnc_UPenn,Winship]{Arif~Rashid}
\author[UPenn,nanded]{Ujjwal~Baid}

%%% who helped with manuscript and baseline implementations
% \author[sustech]{Qingqiao~Hu}

%%%%%%%%%%%%%%%%%%%%%%%%%%%
% participants, re-ordered w.r.t. the last name %
%%%%%%%%%%%%%%%%%%%%%%%%%%%

\author[cairo]{Wesam~Adel}
\author[acibadem]{Deniz~Alis}
\author[nanded]{Bhakti~Baheti}
\author[tmllab]{Yingbin~Bai}
%\author[nanded]{Ujjwal~Baid}
\author[utrecht]{Ishaan~Bhat}
\author[istanbul_tech]{Sabri Can~Cetindag}
\author[Tencent,compsci_shenzen]{Wenting~Chen}
\author[alberta]{Li~Cheng}
\author[nanded]{Prasad~Dutande}
\author[ljubljana]{Lara~Dular}
\author[cairo]{Mustafa A.~Elattar}
\author[tongji]{Ming~Feng}
\author[oppo,beihang]{Shengbo~Gao}
\author[Radboud_dia]{Henkjan~Huisman}
\author[oppo]{Weifeng~Hu}
\author[nanded]{Shubham~Innani}
\author[Tencent,alberta]{Wei~Ji}
\author[harvard_medschool]{Davood~Karimi}
\author[utrecht]{Hugo J.~Kuijf}
\author[korea_uni]{Jin Tae~Kwak}
\author[sejong_uni]{Hoang Long~Le}
\author[harbin]{Xiang~Li}
\author[sustech]{Huiyan~Lin}
\author[tmllab]{Tongliang~Liu}
\author[Nanjing]{Jun~Ma}
\author[Tencent]{Kai~Ma}
\author[harbin_elec,pengcheng_lab,brain_capmed,dis_capmed]{Ting~Ma}
\author[istanbul_tech]{Ilkay~Oksuz}
\author[imperial]{Robbie Holland}
\author[inescid]{Arlindo L.~Oliveira}
\author[ramakrishna]{Jimut Bahan~Pal}
\author[oppo]{Xuan~Pei}
\author[cath_uni,csiro]{Maoying~Qiao}
\author[Radboud_dia]{Anindo~Saha}
\author[copenhagen]{Raghavendra~Selvan}
\author[compsci_shenzen]{Linlin~Shen}
\author[inescid]{Joao Lourenco~Silva}
\author[ljubljana]{Ziga~Spiclin}
\author[nanded]{Sanjay~Talbar}
\author[csiro]{Dadong~Wang}
\author[sustech]{Wei~Wang}
\author[oppo]{Xiong~Wang}
\author[tongji]{Yin~Wang}
\author[sustech]{Ruiling~Xi}
\author[key_lab,def_tech]{Kele~Xu}
\author[harbin_elec]{Yanwu~Yang}
\author[heviAI]{Mert~Yergin}
\author[Tencent]{Shuang~Yu}
\author[sustech]{Lingxi~Zeng}
\author[sustech]{YingLin~Zhang}
\author[hk_scitech]{Jiachen~Zhao}
\author[Tencent]{Yefeng~Zheng}
\author[ljubljana]{Martin~Zukovec}

%%%%%%%%%%%%%%%%%%%%%%%%%%%%%%%%%%
% senior co-organizers / data contributors, to be kept in this ordering %
%%%%%%%%%%%%%%%%%%%%%%%%%%%%%%%%%%%%%

%%% authors to be in the last 

\author[sloan_kettering]{Richard~Do}
\author[sloan_kettering,uzh_radiology]{Anton~Becker}
\author[queen_uni]{Amber~Simpson}
\author[eth]{Ender~Konukoglu}
\author[uzh_kpsi]{Andras~Jakab}
\author[UPenn]{Spyridon Bakas}
\author[Jerusalem]{Leo~Joskowicz}
\author[uzh_dqbm,tum_info]{Bjoern Menze}

%%%%%%%%%%%%%%%%%%%%%%%%%%%%%%%%%%%%%
%%% main addresses %%%
%%%%%%%%%%%%%%%%%%%%%%%%%%%%%%%%%%%%%

\address[tum_info]{Department of Informatics, Technical University of Munich, Germany.}
\address[martinos]{Athinoula A. Martinos Center for Biomedical Imaging, Massachusetts General Hospital, Harvard Medical School, USA.}
\address[uzh_dqbm]{Department of Quantitative Biomedicine, University of Zurich, Switzerland.}
\address[uzh_kpsi]{University Children's Hospital Zurich, University of Zurich, Switzerland.} 
\address[deisar]{Department of Radioncology and Radiation Theraphy , Klinikum rechts der Isar, Technical University of Munich, Germany}
\address[eth]{Department of Information Technology and Electrical Engineering, ETH-Zurich, Switzerland.} 
\address[sloan_kettering]{Department of Radiology, Memorial Sloan Kettering Cancer Center in New York City, USA}
\address[queen_uni]{Department of Biomedical and Molecular Sciences, Queen's University, Canada}
\address[tum_TranslaTUM]{TranslaTUM - Central Institute for Translational Cancer Research, Technical University of Munich, Germany}
\address[MIT]{McGovern Institute, Massachusetts Institute of Technology, USA}
\address[uzh_radiology]{Institute for Diagnostic and Interventional Radiology, Unveristy Zurich Hospital, Switzerland.}
\address[imperial]{BioMedIA, Imperial College London, United Kingdom.}

%%%%%%%%%%%%%%%%%%%%%%%%%%%%%%%%%%%%%
% addresses data contributors %
%%%%%%%%%%%%%%%%%%%%%%%%%%%%%%%%%%%%%

\address[RadOnc_UPenn]{Department of Radiation Oncology, University of Pennsylvania, PA, USA}
\address[UPenn]{University of Pennsylvania, PA, USA}
\address[Winship]{Department of Radiation Oncology, Winship Cancer Institute of Emory University, Georgia, USA}

%%%%%%%%%%%%%%%%%%%%%%%%%%%%%%%%%%%%%
% addresses participants %
%%%%%%%%%%%%%%%%%%%%%%%%%%%%%%%%%%%%%

\address[cairo]{Nile University, Cairo, Egypt}
\address[acibadem]{Department of Medical Sciences, Acibadem University, Istanbul, Turkey}
\address[nanded]{Shri Guru Gobind Singhji Institute of Engineering and Technology, Nanded, Maharashtra, India}
\address[tmllab]{Trustworthy Machine Learning Lab, University of Sydney, Australia}
\address[utrecht]{Image Sciences Institute, University Medical Center Utrecht, The Netherlands}
\address[istanbul_tech]{Computer Engineering Department, Istanbul Technical University, Istanbul, Turkey}
\address[compsci_shenzen]{School of Computer Science, Shenzhen University, Shenzhen, China}
\address[alberta]{University of Alberta, USA}
\address[ljubljana]{University of Ljubljana, Faculty of Electrical Engineering, Ljubljana, Slovenia}
\address[tongji]{Tongji University, Shanghai, China}
\address[oppo]{OPPO Research Institute, Shanghai, China}
\address[beihang]{School of Biological and Medical Engineering, Beihang University, Beijing, China}
\address[harvard_medschool]{Harvard Medical School, Boston, USA}
\address[korea_uni]{School of Electrical Engineering, Korea University, Seoul, Korea}
\address[sejong_uni]{Department of Computer Science and Engineering, Sejong University, Seoul, Korea}
\address[harbin]{Harbin Institute of Technology, China}
\address[sustech]{Southern University of Science and Technology, China}
\address[harbin_elec]{Department of Electronic and Information Engineering, Harbin Institute of Technology at Shenzhen, China}
\address[pengcheng_lab]{Peng Cheng Lab, Shenzhen, China}
\address[brain_capmed]{Advanced Innovation Center for Human Brain Protection, Capital Medical University, Beijing, China}
\address[dis_capmed]{National Clinical Research Center for Geriatric Disorders, Xuanwu Hospital Capital Medical University, Beijing, China}
\address[inescid]{Instituto Superior Tecnico / INESC-ID, Portugal}
\address[ramakrishna]{Department of Computer Science, Ramakrishna Mission Vivekananda Educational and Research Institute, India}
\address[cath_uni]{Australian Catholic University, Australia}
\address[csiro]{Commonwealth Scientific and Industrial Research Organisation (CSIRO), Australia}
\address[copenhagen]{University of Copenhagen, Denmark}
\address[key_lab]{National Key Lab of Parallel and Distributed Processing, Changsha, China}
\address[def_tech]{National University of Defense Technology, Changsha, China}
\address[heviAI]{Hevi AI, Istanbul, Turkey}
\address[hk_scitech]{Department of Computer Science and Engineering, Hongkong University of Science and Technology, China}
\address[Tencent]{Tencent Healthcare (Shenzhen) Co., Ltd, China}
\address[Nanjing]{Department of Mathematics, Nanjing University of Science and Technology, China}
\address[Radboud_dia]{Diagnostic Image Analysis Group, Radboud University Medical Center, Nijmegen, The Netherlands}
\address[Jerusalem]{The Rachel and Selim Benin School of Computer Science and Engineering, The Hebrew University of Jerusalem, Israel}
\address[hai]{Helmholtz AI, Helmholtz Zentrum München, Germany}
\address[translatum]{TranslaTUM - Central Institute for Translational Cancer Research, Technical University of Munich, Germany}
\address[tum_neuro]{Department of Diagnostic and Interventional Neuroradiology, School of Medicine, Klinikum rechts der Isar, Technical University of Munich, Germany}